\documentclass[12pt,preprint]{aastex}
\shorttitle{DETECTION OF ACETONE ON ORION-KL}
\shortauthors{Friedel et al.}
\begin{document}
\newcommand\acetone{(CH$_3$)$_2$CO}
\newcommand\dme{(CH$_3$)$_2$O}
\newcommand\mef{HCOOCH$_3$}
\newcommand\fa{HCOOH}
\newcommand\vycn{C$_2$H$_3$CN}
\newcommand\etcn{C$_2$H$_5$CN}
\newcommand\etoh{C$_2$H$_5$OH}
\newcommand\nt{$2.0(0.3)\times10^{16}$~cm$^{-2}$}
\newcommand\trot{176(48)~K}
\newcommand\nts{$8.0(1.2)\times10^{16}$~cm$^{-2}$}
\newcommand\trots{194(66)~K}
\newcommand\ntr{($2.0(0.3)-8.0(1.2))\times10^{16}$~cm$^{-2}$}
\newcommand\tr{176(48)-194(66)~K}

\title{DETECTION OF INTERSTELLAR ACETONE TOWARD THE ORION-KL HOT CORE}

\author{D. N. Friedel\altaffilmark{1}, L. E. Snyder\altaffilmark{1}, Anthony J. Remijan\altaffilmark{1,2,3}, and B. E. Turner\altaffilmark{4}}

\altaffiltext{1}{Department of Astronomy, 1002 W. Green St., University of
Illinois, Urbana IL 61801\\
email: friedel@astro.uiuc.edu, snyder@astro.uiuc.edu}
\altaffiltext{2}{NASA Goddard Space Flight Center, Computational and Information Sciences and Technologies Office, Code 606
, Greenbelt, MD 20771\\
email: aremijan@pop900.gsfc.nasa.gov}
\altaffiltext{3}{National Research Council Resident Research Associate}
\altaffiltext{4}{National Radio Astronomy Observatory, Charlottesville, VA 22903\\email: bturner@nrao.edu}

\begin{abstract}
We present the first detection of interstellar acetone [\acetone] toward the high mass star forming region
Orion-KL and the first detection of vibrationally excited \acetone\ in the ISM. 
Using the BIMA Array, 28 emission features that can be assigned to 54 acetone transitions were detected. 
Furthermore, 37 of these transitions have not been previously observed in the ISM. The 
observations also show that the acetone emission is concentrated toward the hot core region of Orion-KL,
contrary to the distribution of other large oxygen bearing molecules. From our rotational-temperature diagram
we find a beam averaged \acetone\
column density of \ntr\ and a rotational temperature of \tr.
\end{abstract}

\keywords{ISM:individual(Orion-KL)---ISM:molecules---radio lines:ISM}

The first detection of interstellar acetone [\acetone] was reported by \citet{combes87} and later confirmed by \citet{snyder02}
towards the hot molecular core source Sgr~B2(N-LMH). Previous to this paper, acetone had yet to be reported in any other high 
mass or low mass star forming region despite being structurally similar to dimethyl ether (\dme).  Conversely, \dme\ has been 
detected in both low mass star forming regions including IRAS 16293-2422 \citep{cazaux03} and high mass star 
forming regions including Sgr \citep{winn76} and Orion \citep{snyder74}.  It seemed therefore likely that acetone may be prevalent in
other star forming regions such as Orion-KL.

The Orion-KL region is the closest ($\sim$480 pc) site of massive star formation \citep{genzel81}. 
There are several cloud components (e.g. hot core, compact ridge, extended ridge, and plateau) which are associated 
with Orion-KL, each with different chemical and physical properties \citep[e.g.][]{blake87}. The two most chemically interesting components,
the hot core and compact ridge, are separated by only $\sim$5800 AU. Large oxygen bearing species (such as methyl formate (\mef), 
dimethyl ether (\dme), and formic acid (\fa)) are observed primarily toward the compact ridge
\citep[e.g.][]{liu02}, while large nitrogen bearing species (i.e.\ vinyl cyanide (\vycn) and ethyl cyanide (\etcn)) 
are located toward the hot core \citep[e.g.][]{blake87}.

In this letter, we present the first detection of interstellar \acetone\ toward the high mass star forming region 
Orion-KL and the first detection of vibrationally excited \acetone\ in the ISM. Using the Berkeley-Illinois-Maryland Association (BIMA)\footnote{Operated by the
University of California, Berkeley, the University of Illinois, and the
University of Maryland with support from the National Science Foundation.} 
Array, 28 emission features that can be assigned to 54 acetone transitions 
were detected toward the Orion hot core. 
The detection of \acetone\ toward the Orion hot core is unusual since it is a complex oxygen bearing species, structurally similar 
to \dme, a compact ridge species. This indicates that the formation of \acetone\ may be very different than \dme\ and may invoke 
N bearing species or grain surface reactions.  Finally, since acetone is prevalent toward the Orion hot core region, it is most
likely that a spectral line assigned as interstellar glycine (NH$_2$CH$_2$COOH) toward the Orion-KL region can be assigned instead to interstellar 
\acetone.

The data presented were taken between 2002 and 2003 November as part of a 3mm spectral line survey of the Orion-KL region by the 
BIMA Array (Friedel et al., 2006 in prep.) in its C configuration. 
The phase center of our observations was 
$\alpha$(J2000)=$05^h35^m4^s.3$, $\delta$(J2000)=$-05{\degr}22{\arcmin}27{\farcs}6$, which is near the hot core 
($\alpha$(J2000)=$05^h35^m14^s.5$, $\delta$(J2000)=$-5{\degr}22{\arcmin}30{\arcsec}$) and the compact ridge 
($\alpha$(J2000)=$05^h35^m14^s.2$, $\delta$(J2000)=$-5{\degr}22{\arcmin}41{\arcsec}$). The array was operating in cross-correlation 
(double-sideband) mode with a sideband rejection of better than 20 dB. The correlator was configured to have four 50 MHz wide windows in each sideband 
with 128 channels per window. This resulted in a channel spacing of 391 kHz ($\sim1.4-1.1$ km s$^{-1}$), for all 
observations. Saturn was used as the flux density calibrator and 0423-013 and 0609-157 were used to calibrate the antenna based gains. 
The absolute amplitude calibration of 0423-013 and 0609-157 from the flux density calibrators is accurate to within $\sim$20\%. The 
passbands were automatically calibrated during data acquisition. 
The data were calibrated, continuum subtracted and imaged using the MIRIAD software package \citep{sault95}. 
The ground state \acetone\ molecular parameters were taken from \citet{groner02} and the vibrational \acetone\ molecular
parameters were taken from \citet{groner05}.

Table~\ref{tab:acetone} lists the molecular parameters and observational results for the detected acetone lines. 
Column 1 lists the rest frequency in MHz, column 2 lists the quantum numbers, column 3 lists the upper state energy ($E_u$) in K, column 4 lists the line strength ($S_{ij}$), column 5 lists 
the spin weight, column 6 lists the integrated line intensity ($W$) in Jy beam$^{-1}$ km s$^{-1}$, column 7 lists the rest velocity of the transitions ($V_{LSR}$) in
km s$^{-1}$, column 8 lists the synthesized beam 
size in arcseconds, and column 9 lists the figure in which the associated spectra appear. 
We detected a total of 28 spectral features that 
can be assigned to 54 \acetone\ transitions, 37 of which have not been previously observed in the ISM, toward the Orion hot core. Of these 54 transitions 46 are from the
ground state and 8 are from the first vibrationally excited state which lies at 115 K above ground. No
\acetone\ transitions were seen, above our 3 $\sigma$ limit, toward the Orion compact ridge.

\begin{figure}
\plotone{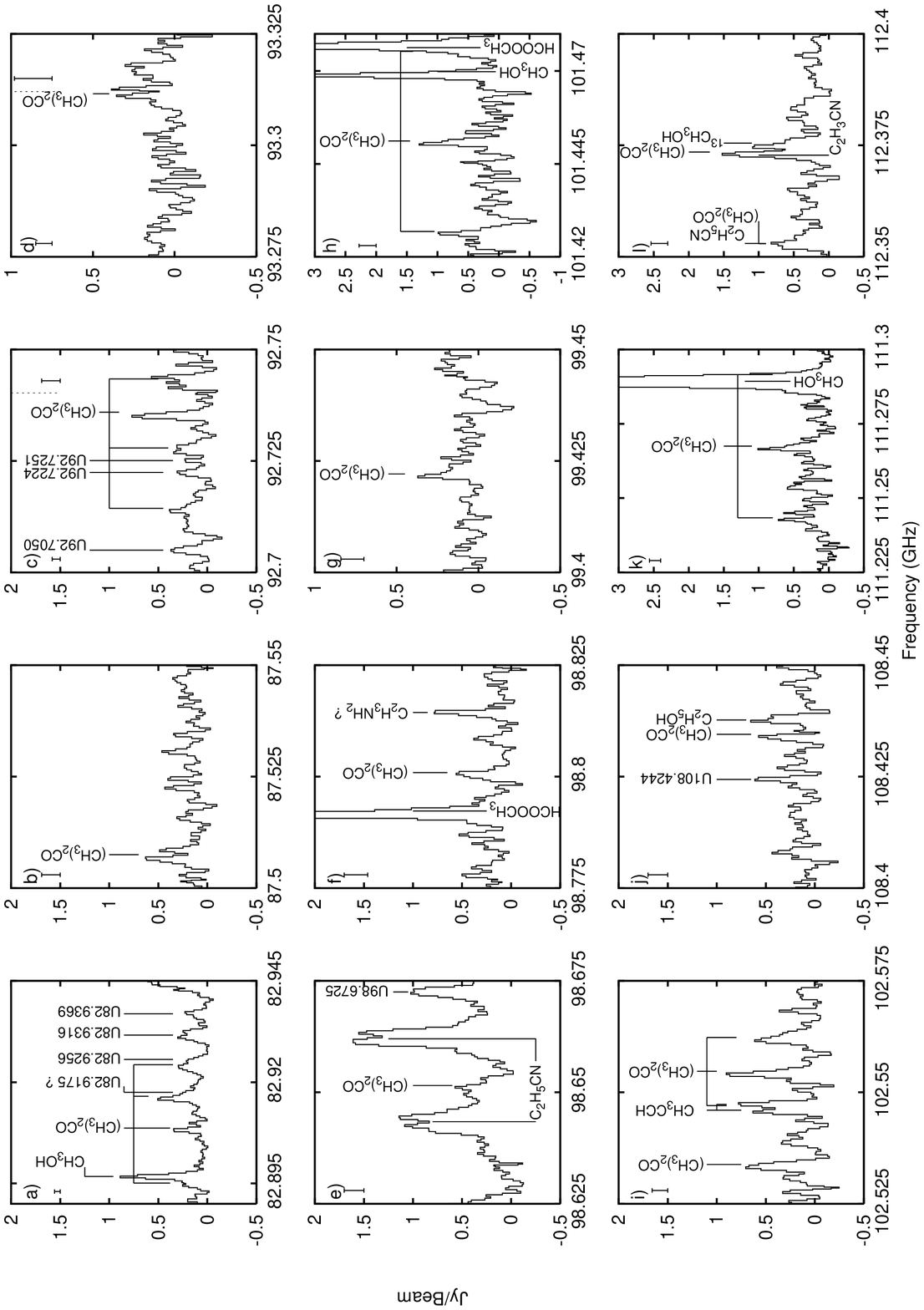}
\caption{Spectra of each of the detected acetone (\acetone) transitions toward the Orion hot core, along with nearby 
identified and unidentified transitions.  All spectra have been hanning smoothed over 3 channels.
The ``I'' bar in the upper left corner of each panel denotes the 1 $\sigma$ rms noise level for the respective panel. The dashed line
and second ``I'' bar in panel c) indicates where a second window is located and its associated 1 $\sigma$ rms noise level.\label{fig:spec}}
\end{figure}

Figure~\ref{fig:spec} shows the spectra of each of our detected transitions and nearby identified and unidentified lines. All spectra are 
hanning smoothed over 3 channels and the spectral line fiducials are for a $V_{LSR}$ of 5 km s$^{-1}$. The ``I'' bar in the upper left corner of each plot denotes the 
1 $\sigma$ rms noise level for the plot. 
In subplots c) and d) the dashed line and second ``I'' bar denote where a second spectral window was located and the associated 1 $\sigma$ rms noise level 
of that window.

\begin{figure}
\plotone{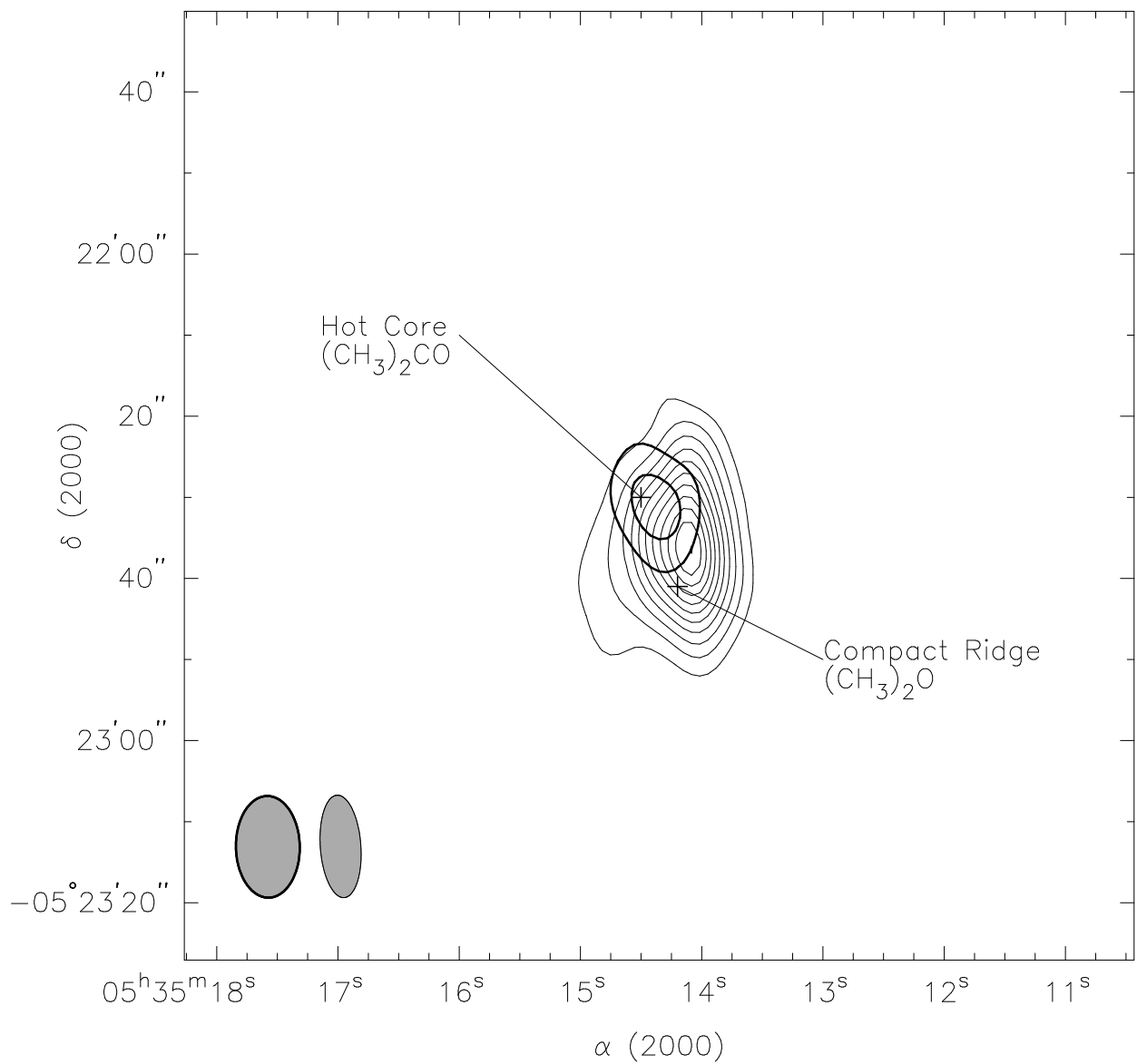}
\caption{Contour maps of acetone and dimethyl ether toward the Orion KL region. Heavy contours are 
the $8_{*,8}-7_{*,7} EE$ degenerate transitions of \acetone\ ($E_u$=19 K). Contour levels
are $\pm3,5$ $\sigma$ ($\sigma=0.09$ Jy beam$^{-1}$). The normal contours are the
$7_{0,7}-6_{1,6}$ degenerate transitions of \dme\ ($E_u$=25 K, $\nu$=111.783112 GHz). Contour spacings are 4 $\sigma$ starting at 
4 $\sigma$ ($\sigma=0.24$ Jy beam$^{-1}$).  The hot core and compact ridge are labeled with ``+''. The synthesized beams
for the respective maps are in the lower left corner.\label{fig:map}}
\end{figure}

Figure~\ref{fig:map} shows the map of the $8_{*,8}-7_{*,7} EE$ degenerate transitions of \acetone\ ($E_u$=19 K) in heavy contours. Contours are $\pm$3 and 5 $\sigma$
($\sigma$=0.09 Jy beam$^{-1}$). The normal contours are the $7_{0,7}-6_{1,6}$ four fold degenerate transition of \dme\ ($E_u$=25 K, $\nu$=111.783112 GHz). Contour spacing is 4 $\sigma$
starting at 4 $\sigma$ ($\sigma$=0.24 Jy beam$^{-1}$).
The hot core and compact ridge are labeled and indicated by the ``+'' signs. 
Note that the \acetone\ emission traces closer to the hot core while the \dme\ emission traces closer to the compact ridge. The synthesized beams for each species are in the lower left
corner of the map. The average $V_{LSR}$ for the detected \acetone\ transitions is 6.4(1.3)\footnote{All uncertainties are 1 $\sigma$ unless otherwise noted.} km s$^{-1}$ which is between the $V_{LSR}$ for the hot core ($\sim$5 km s$^{-1}$) and
the compact ridge ($\sim$8 km s$^{-1}$). Yet it is consistent with the $V_{LSR}$ of other molecular species \citep{blake87}. The maps of the transitions, however,
 indicate that \acetone\ is more closely associated with the hot core rather than
the compact ridge \citep{blake87}.

\begin{figure}
\plotone{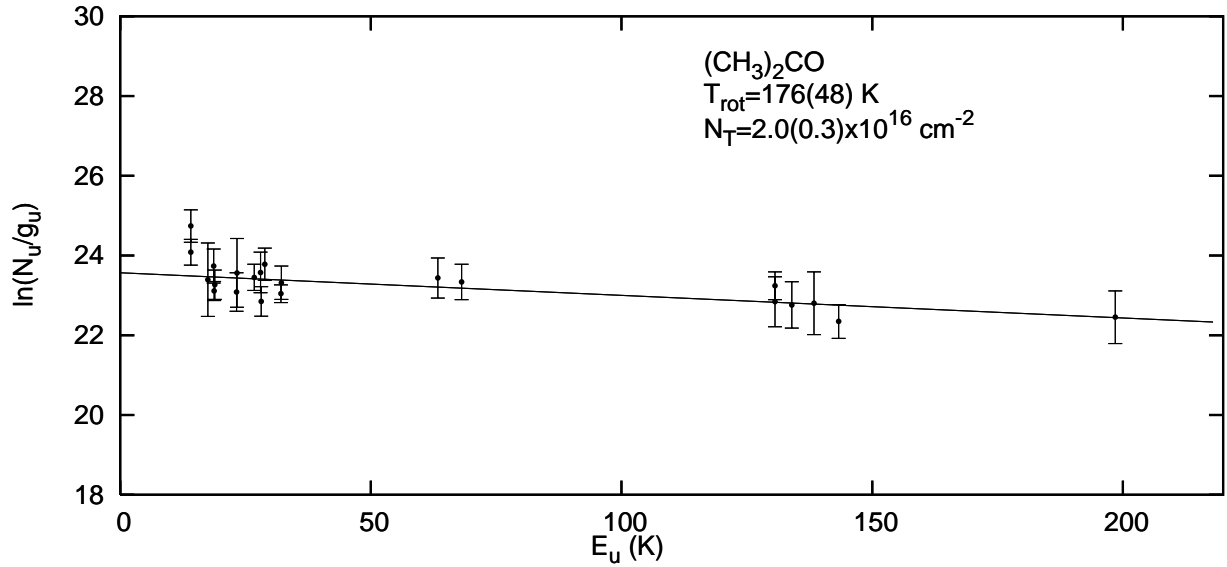}
\caption{Rotational temperature diagram for detected acetone transitions. The ordinate is $ln(\langle N_u\rangle/g_u)$, the abscissa is $E_u$ in K, and the uncertainties are 1 $\sigma$. The
weighted least squares fit is shown by the line. The resulting rotational temperature and column density are shown.\label{fig:rtd}}
\end{figure}

For interferometric observations the total column density $\langle N_T\rangle$ can be calculated from \citep{snyder02,snyder05}:
\begin{equation}
\frac{\langle N_T\rangle}{q_{rv}}e^{-E_u/T_{rot}}=\frac{\langle N_u\rangle}{g_u}=\frac{2.04}{B\theta_a\theta_bS\mu^2\nu^3}\left(\frac{W}{g}\right)\times10^{20} cm^{-2}
\end{equation}
where $\langle N_u\rangle/g_u$ is the beam averaged upper level column density over the statistical weight, $B$ is the beam filling factor, $g$ is the spin weight, 
 $\mu$ is the dipole moment ($\mu_b$=2.93 D), $\nu$ is the transition frequency in GHz, $T_{rot}$ is the 
rotational temperature and $E_u$ is the upper level energy of the transition in K. The rotational-vibrational partition function, $q_{rv}$, is defined as
\begin{equation}
q_{rv}\approx\sum_{i=0}^{2}e^{-E_i/T_{rot}}q_r,
\end{equation}
where $E_i$ are the energies above ground for the ground state and the first two vibrational states (0 K, 115 K, and 180 K, respectively) and $q_r$ is the rotational partition function\footnote{The rotational part of the partition function is approximated by the ground state $q_r$ since $q_r$ for the first vibrational state differs from the ground state by less than 1 part in 1000 and $q_r$ for the second vibrational state has not been reported.}
(261.7$T_{rot}^{3/2}$)\citep{groner02,groner05,ww05}. 
Even though we did not detect any
transitions from the second vibrationally excited state, it lies only 180 K above ground and thus will have a significant population at temperatures above 100 K.
For a set of degenerate transitions, $g$ is the sum of the spin weights of the blended states.
By plotting $ln(\langle N_u\rangle/g_u)$, for each integrated line intensity in Table~\ref{tab:acetone}, versus $E_u$ we can obtain the total beam averaged column density
and rotational temperature of \acetone. 
Figure~\ref{fig:rtd} shows the rotational-temperature diagram (assuming the source fills our synthesized beams, $B=0.5$, as discussed by \citet{snyder05}) and
yields $\langle N_T\rangle=$\nt\ and $T_{rot}=$\trot. If instead we assume a source size similar to the hot core continuum size seen by \citet{liu02} of 5\arcsec$\times$3\arcsec, the diagram yields $\langle N_T\rangle=$\nts\ and $T_{rot}=$\trots. Both temperatures are reasonable considering the range in temperatures observed toward the hot core 
($\sim$100 K \citep[\etcn,][]{white03} to $\sim$300 K \citep[HC$_3$N,][]{wright96}). From these column densities, rotation temperatures, and \citet{turner91} we calculated the opacity to be $<0.02$ for all transitions.
 For a H$_2$ column density of $\sim5\times10^{24}$ cm$^{-2}$ \citep[][and references therein]{kaufman98} 
we find a fractional abundance of $(4.0(0.6)-16.0(2.4))\times10^{-9}$ which is higher than the fractional abundance of \acetone\ toward Sgr~B2(N-LMH) ($4-30\times10^{-10}$) 
\citep{snyder02}. The difference however can be attributed our use of the vibrational part of the partition function. If we recalculate
the abundances from Sgr~B2(N-LMH) including the vibrational part of the partition function it agrees with our abundance toward Orion.

The detection of \acetone\ toward the hot core and the non-detection (above 3 $\sigma$) toward the compact ridge,
makes \acetone\ the only known highly saturated O species that is seen
only toward the hot core. This suggests a formation mechanism such as high temperature gas phase reactions (i.e. neutral-neutral) or grain surface reactions,
as was suggested by \citet{herbst90}, rather than formation in or behind a shock front (i.e. ion-molecule chemistry or liberation from the grain surface).

Finally, the detection of acetone toward the Orion hot core is also very significant because it shows that a spectral feature
previously identified as glycine may in fact be due to interstellar acetone. \citet{kuan03} noted that one of the lines assigned to glycine 
was coincident with the $31_{*,19}-31_{*,20} EE$ degenerate \acetone\ pair. The identification of a transition of
\acetone\ was dismissed as unlikely due to its unfavorable quantum numbers. However, we can calculate the expected 
integrated intensity for these degenerate acetone transitions using the \acetone\ column densities from this
work and from eqs. [1] \& [2] of \citet{snyder05}. Solving eqs. [1] \& [2] of \citet{snyder05} for the integrated line
intensity ($W$) we find:
\begin{equation}
W=\frac{\langle N_T\rangle gS\mu^2{\nu}B^*}{1.67q_{rv}e^{E_u/T_{rot}}}\times10^{-14}~K~km~s^{-1},
\end{equation}
where, for these transitions, $g=16+16=32$, $S=12.555$, $\nu=164.870$ GHz, and $E_u=364$ K. Also $B^*=\eta^*_MB$, where $\eta^*_M$ is the corrected 
main beam efficiency for the 12 meter ($\eta^*_M\sim0.7$ at 164 GHz). If we use the 5\arcsec$\times$3\arcsec\ source size and associated $\langle N_T\rangle$ and $T_{rot}$ we calculate an integrated line intensity of 0.218(0.031)~K~km~s$^{-1}$. Comparing this value to that reported by \citet{kuan03} of 0.091(0.035)~K~km~s$^{-1}$ we see that the integrated line intensity overlaps with the reported value at the 2 $\sigma$ level.
Because of the detection of 54 transitions of interstellar \acetone\ toward
the Orion hot core, it is most likely that the transition attributed to glycine by Kuan et al.\ (2003) is due to \acetone.

We have reported the detection of 54 \acetone\ transitions toward the Orion-KL hot core region with a column density of 
\ntr\ and a rotational temperature of \tr. We also have reported the first detection of vibrationally exited \acetone\ in the ISM. The detection of a large O bearing species purely toward the hot core suggests 
that the formation of \acetone\ may be very different than \dme\ and may invoke 
N bearing species or grain surface reactions.

\acknowledgements
We thank P. Groner and B. Drouin for supplying useful data. We also thank an anonymous referee for a very favorable review of this work. This work was partially funded by grant NSF AST02-28953 and the University of Illinois.

\clearpage

\begin{deluxetable}{ccrrccccc}
\tabletypesize{\scriptsize}
\tablecolumns{9}
\tablecaption{Acetone (\acetone) Molecular Line and Observational Parameters\tablenotemark{a}}
\tablehead{
\colhead{} & \colhead{} & \colhead{} & \colhead{} & \colhead{} & \colhead{$W$} & \colhead{} & \colhead{} & \colhead{}\\
\colhead{Frequency} & \colhead{Quantum} & \colhead{$E_{u}$} & \colhead{} & \colhead{Spin} & \colhead{(Jy bm$^{-1}$} & \colhead{$V_{LSR}$} & \colhead{$\theta_a\times\theta_b$} & \colhead{}\\
\colhead{(MHz)\tablenotemark{b}} & \colhead{Numbers} & \colhead{(K)} & \colhead{$S_{ij}$} & \colhead{Wt.} & \colhead{km s$^{-1}$)} & \colhead{(km s$^{-1}$)} & \colhead{($\arcsec$)} & \colhead{Fig.}
}
\startdata
 81,813.725 (12) & $7_{1,6}-6_{2,5} EE $  & 17.47  & 5.21 & 16 & 1.5(1.0) & 6.6(2.5) & $12.99\times8.33$ & \tablenotemark{c}\\
 82,895.112 (52) & $8_{*,8}-7_{*,7} EE, \nu=1$\tablenotemark{d,e} & 133.95 & 7.38 & 32 & 2.1(0.6) & 6.0(2.3) & $12.54\times7.86$ & 1a\\
 82,908.654 (20)& $8_{*,8}-7_{*,7} AE $  & 18.85  & 7.38 & 8 & 1.7(0.3) & 6.9(0.9) & $12.54\times7.86$ & 1a\\
 82,908.702 (20)& $8_{*,8}-7_{*,7} EA $  & 18.85  & 7.38 & 8 & \tablenotemark{f} &  & & 1a\\
 82,916.525 (14) & $8_{*,8}-7_{*,7} EE $  & 18.74  & 7.38 & 32 & 3.0(0.3) & 7.3(0.4) & $12.54\times7.86$ & 1a\\
 82,924.325 (22)& $8_{*,8}-7_{*,7} AA $  & 18.64  & 7.38 & 16 & 2.8(0.5) & 7.5(1.1) & $12.54\times7.86$ & 1a\\
 87,507.547 (46)& $18_{8,10}-18_{7,11} EE$\tablenotemark{d,e} & 130.60 & 8.27 & 16 & 2.8(0.3) & 6.6(2.4) & $13.16\times9.21$ & 1b\\
 87,580.004 (44)& $18_{9,10}-18_{8,11} EE$\tablenotemark{d,e} & 130.60 & 8.27 & 16 & 1.4(0.7) & 4.5(0.5) & $11.82\times7.81$ & \tablenotemark{c}\\
 92,714.466 (54)& $9_{*,9}-8_{*,8} EE, \nu=1$\tablenotemark{d,e} & 138.41 & 8.38 & 32 & 3.1(0.8) & 8.3(6.9) & $13.89\times6.20$ & 1c\\
 92,727.906 (22)& $9_{*,9}-8_{*,8} AE $  & 23.30  & 8.38 & 8 & 3.3(0.9) & 5.8(0.8) & $13.89\times6.20$ & 1c\\
 92,727.952 (20)& $9_{*,9}-8_{*,8} EA $  & 23.30  & 8.38 & 8 & \tablenotemark{f} & & & 1c\\
 92,735.672 (16) & $9_{*,9}-8_{*,8} EE $  & 23.19  & 8.38 & 32 & 4.1(0.5) & 6.9(1.9) & $13.89\times6.20$ & 1c\\
 92,743.363 (24) & $9_{*,9}-8_{*,8} AA $  & 23.09  & 8.38 & 16 & \tablenotemark{g} &  & & 1c\\
 93,311.552 (90) & $22_{12,11}-22_{11,12} EE$\tablenotemark{d,e} & 198.48 & 11.29 & 16 & 1.8(0.7) & 6.5(1.7) & $14.86\times6.91$ & 1d\\
 98.651.514 (20) & $5_{5,1}-4_{4,1} EE$\tablenotemark{d,e} & 14.09 & 3.63 & 16 & 5.6(0.4) & 7.0(3.0) & $13.86\times6.19$ & 1e\\
 98,800.980 (14) & $5_{5,0}-4_{4,0} EE$\tablenotemark{d,e} & 14.09 & 3.63 & 16 & 3.7(0.3) & 6.2(1.3) & $15.54\times7.03$ & 1f\\
 99,422.080 (24) & $14_{*,11}-14_{*,12} EE$\tablenotemark{d,e} & 68.11 & 3.33 & 32 & 2.7(0.5) & 4.7(2.1) & $13.64\times6.61$ & 1g\\
101,426.716 (19) & $9_{1,8}-8_{2,7} AE/EA\tablenotemark{d} $  & 26.84  & 7.21 & 10 & 4.7(0.4) & 6.3(0.5) & $12.91\times6.74$ & 1h\\
101,427.090 (19) & $9_{2,8}-8_{1,7} AE/EA\tablenotemark{d} $  & 26.84  & 7.21 & 6 & \tablenotemark{f} & & & 1h\\
101,451.059 (14) & $9_{1,8}-8_{2,7} EE\tablenotemark{d} $  & 26.74  & 7.21 & 16 & 7.2(0.3) & 6.7(1.0) & $13.35\times6.94$ & 1h\\
101,451.446 (14) & $9_{2,8}-8_{1,7} EE\tablenotemark{d} $  & 26.74  & 7.21 & 16 & \tablenotemark{f} & & & 1h\\
101,475.332 (22) & $9_{1,8}-8_{2,7} AA\tablenotemark{d} $  & 26.64  & 7.21 & 10 & \tablenotemark{h} & & & 1h\\
101,475.733 (22) & $9_{2,8}-8_{1,7} AA\tablenotemark{d} $  & 26.64  & 7.21 & 6 & \tablenotemark{h} & & & 1h\\
102,533.756 (58) &$10_{*,10}-9_{*,9} EE, \nu=1$\tablenotemark{d,e,i}&~~143.33 & 9.37 & 32 & 2.7(0.5) & 6.3(1.5) & $12.80\times6.17$ & 1i\\
102,547.058 (22) & $10_{0,10}-9_{1,9} AE/EA\tablenotemark{d} $ & 28.22  & 9.38 & 16 & \tablenotemark{j} & & & 1i\\
102,554.696 (18) & $10_{0,10}-9_{1,9} EE\tablenotemark{d} $ & 28.11  & 9.38 & 32 & 4.5(0.4) & 6.9(1.9) & $12.80\times6.17$ & 1i\\
102,562.281 (26) & $10_{0,10}-9_{1,9} AA\tablenotemark{d} $ & 28.01  & 9.38 & 16 & 4.7(0.5) & 6.8(2.4) & $12.80\times6.17$ & 1i\\
108,434.511 (32) & $14_{2,12}-14_{1,13} EE$\tablenotemark{d,e} & 63.34 & 2.29 & 32 & 2.0(0.5) & 6.9(1.1) & $12.22\times5.64$ & 1j\\
111,243.359 (22)& $10_{*,9}-9_{*,8} AE\tablenotemark{d} $ & 32.17  & 8.21 & 8 & 3.6(0.4) & 7.1(1.2) & $12.44\times5.77$ & 1k\\
111,243.448 (20)& $10_{*,9}-9_{*,8} EA\tablenotemark{d} $ & 32.17  & 8.21 & 8 & \tablenotemark{f} &  & & 1k\\
111,267.540 (16) & $10_{*,9}-9_{*,8} EE\tablenotemark{d} $ & 32.07  & 8.21 & 32 & 5.6(0.2) & 6.6(0.9) & $12.44\times5.77$ & 1k\\
111,291.600 (24) & $10_{*,9}-9_{*,8} AA\tablenotemark{d} $ & 31.98  & 8.21 & 16 & \tablenotemark{k} &  & & 1k\\
112,352.908 (60) & $11_{*,11}-10_{*,10} EE, \nu=1$\tablenotemark{d,e} & 148.71 & 10.37 & 32 & \tablenotemark{l}&  & & 1l\\
112,373.548 (18) & $11_{0,11}-10_{1,10} EE$\tablenotemark{d,e} & 33.51 & 10.39 & 32 & \tablenotemark{m} &  & & 1l\\

\enddata
\tablenotetext{a}{Degenerate pairs of transitions are denoted with a * in the quantum numbers or list two torsional states (i.e. $AE/EA$). The reported frequency is the average frequency of the transitions and the reported spin weight is the sum from the transitions. See \citet{groner02} and \citet{groner05} for a complete listing.}
\tablenotetext{b}{Frequency uncertainty is 2 $\sigma$.}
\tablenotetext{c}{This transition will be shown in Friedel et al. (2006, in prep.).}
\tablenotetext{d}{Not previously observed in ISM.}
\tablenotetext{e}{The $AA$, $AE$ and $EA$ transitions were below the 3 $\sigma$ detection limit.}
\tablenotetext{f}{This transition is blended with the previous one(s) and the integrated line intensity and column density are calculated as such.}
\tablenotetext{g}{The $AA$ transitions are less than 3 $\sigma$ and thus are not used in the calculations.}
\tablenotetext{h}{This transition is blended with the $18_{3,15}-18_{3,16} E$ transition of methyl formate (\mef).}
\tablenotetext{i}{This transition of \acetone\ is coincident with the $6_{0,6}-5_{0,5}$ transition of ethanol (\etoh), however since \etoh\ is thought to be a compact ridge species \citep{schilke97} and is only weakly detected toward the compact ridge, we consider the contamination from \etoh\ to be minimal.}
\tablenotetext{j}{This transition is blended with the $6_0-5_0$ transition of propyne (CH$_3$CCH).}
\tablenotetext{k}{This transition is blended with the $7_{2, 5}-8_{1, 8} A+$ transition of methanol (CH$_3$OH).}
\tablenotetext{l}{This transition is blended with the $11_{1,11}-10_{0,10}A$ $\nu_t=1$ transition of ethyl cyanide (\etcn).}
\tablenotetext{m}{This transition may be blended with the $8_{1,8}-7_{0,7}$ $^1\nu_{11}$ transition of vinyl cyanide (\vycn) and is not used in the calculations.}
\label{tab:acetone}
\end{deluxetable}

\clearpage

\clearpage

\clearpage

\clearpage

\begin{thebibliography}{}
\bibitem[Blake et al.(1987)]{blake87} Blake, G.~A., Sutton, E.~C., Masson, C.~R., \& Phillips, T.~G.\ 1987, \apj, 315, 621 

\bibitem[Cazaux et al.(2003)]{cazaux03} Cazaux, S., Tielens, A.~G.~G.~M., Ceccarelli, C., Castets, A., Wakelam, V., Caux, E., Parise, B., \& Teyssier, D.\ 2003, \apjl, 593, L51 
 
\bibitem[Combes et al.(1987)]{combes87} Combes, F., Gerin, M., Wootten, A., Wlodarczak, G., Clausset, F., \& Encrenaz, P.~J.\ 1987, \aap, 180, L13 
 
\bibitem[Kuan et al.(2003)]{kuan03} Kuan, Y., Charnley, S.~B., Huang, H., Tseng, W., \& Kisiel, Z.\ 2003, \apj, 593, 848 

\bibitem[Genzel et al.(1981)]{genzel81} Genzel, R., Reid, M.~J., Moran, J.~M., \& Downes, D.\ 1981, \apj, 244, 884 

\bibitem[Groner et al.(2005)]{groner05} Groner, P., Herbst, E., De Lucia, F.~C., Drouin, B.~J., \& M\"{a}der, H. 2005, 60th OSU International Conference on Molecular Spectroscopy, RA03, 206

\bibitem[Groner et al.(2002)]{groner02} Groner, P., Albert, S., Herbst, E., De Lucia, F.~C., Lovas, F.~J., Drouin, B.~J., \& Pearson, J.~C.\ 2002, \apjs, 142, 145  

\bibitem[Herbst et al.(1990)]{herbst90} Herbst, E., Giles, K., \& Smith, D.\ 1990, \apj, 358, 468 
 
\bibitem[Kaufman et al.(1998)]{kaufman98} Kaufman, M.~J., Hollenbach, D.~J., \& Tielens, A.~G.~G.~M.\ 1998, \apj, 497, 276 
 
\bibitem[Liu et al.(2002)]{liu02} Liu, S., Girart, J.~M., Remijan, A., \& Snyder, L.~E.\ 2002, \apj, 576, 255  

\bibitem[Sault et al.(1995)]{sault95} Sault, R.~J., Teuben, P.~J., \& Wright, M.~C.~H.\ 1995, ASP Conf.~Ser.~ 77: Astronomical Data Analysis Software and Systems IV, 77, 433 

\bibitem[Schilke et al.(1997)]{schilke97} Schilke, P., Groesbeck, T.~D., Blake, G.~A., \& Phillips, T.~G.\ 1997, \apjs, 108, 301 

\bibitem[Snyder et al.(1974)]{snyder74} Snyder, L.~E., Buhl, D., Schwartz, P.~R., Clark, F.~O., Johnson, D.~R., Lovas, F.~J., \& Giguere, P.~T.\ 1974, \apjl, 191, L79 

\bibitem[Snyder et al.(2002)]{snyder02} Snyder, L.~E., Lovas, F.~J., Mehringer, D.~M., Miao, N.~Y., Kuan, Y., Hollis, J.~M., \& Jewell, P.~R.\ 2002, \apj, 578, 245 
 
\bibitem[Snyder et al.(2005)]{snyder05} Snyder, L.~E., et al.\ 2005, \apj, 619, 914 

\bibitem[Turner(1991)]{turner91} Turner, B.~E.\ 1991, \apjs, 76, 617 
 
\bibitem[Ulich \& Haas(1976)]{ulich76} Ulich, B.~L., \& Haas, R.~W.\ 1976, \apjs, 30, 247 

\bibitem[White et al.(2003)]{white03} White, G.~J., Araki, M., Greaves, J.~S., Ohishi, M., \& Higginbottom, N.~S.\ 2003, \aap, 407, 589 

\bibitem[Widicus Weaver et al.(2005)]{ww05} Widicus Weaver, S.~L., Butler, R.~A.~H., Drouin, B.~J., Petkie, D.~T., Dyl, K.~A., De Lucia, F.~C., \& Blake, G.~A.\ 2005, \apjs, 158, 188 
 
\bibitem[Winnewisser \& Gardner(1976)]{winn76} Winnewisser, G., \& Gardner, F.~F.\ 1976, \aap, 48, 159 
 
\bibitem[Wright et al.(1996)]{wright96} Wright, M.~C.~H., Plambeck, R.~L., \& Wilner, D.~J.\ 1996, \apj, 469, 216 

\end{thebibliography}
\end{document}